\documentclass[aps,pre,twocolumn,groupedaddress,showpacs,amsmath,amssymb]{revtex4-1}

\usepackage{graphicx,color}
\usepackage{epstopdf}

\usepackage{bm}
\newcommand{\trpc}[1]{{{#1}}}
\newcommand{\trpa}[1]{{{#1}}}
\newcommand{\trp}[1]{{{#1}}}
\newcommand{\hs}[1]{{{#1}}}
\newcommand{\hsn}[1]{{{#1}}}

\begin{document}

\def\ldotsplus{\mathinner{\ldotp\ldotp\ldotp\ldotp}}
\def\fourdots{\relax\ifmmode\ldotsplus\else$\m@th \ldotsplus\,$\fi}


\title{Wrinkling of a thin \trpa{film} on a nematic liquid crystal elastomer}

\author{Harsh Soni$^1$}
\author{Robert A. Pelcovits$^2$}
\author{Thomas R. Powers$^{1,2}$}

\affiliation{$^1$School of Engineering, Brown University, Providence, RI 02912 USA}
\affiliation{$^2$Department of Physics, Brown University, Providence, RI 02012 USA}

\date{April 30, 2016}

\begin{abstract}
 \hsn{Wrinkles \trpa{commonly} develop in a thin film deposited on 
 \trpa{a soft} elastomer substrate \trpa{when the film is subject to compression.} 
 \trpa{Motivated by recent experiments [Agrawal et al., Soft Matter \textbf{8}, 7138 (2012)] that show how wrinkle morphology can be controlled by using a nematic elastomer substrate, we develop the theory of small-amplitude wrinkles of an isotropic film atop a nematic elastomer. The  directors of the nematic elastomer are assumed to lie in a plane parallel to the plane of the undeformed film. For uniaxial compression of the film along the direction perpendicular to the elastomer directors,  the system behaves as a compressed film on an isotropic substrate. When the uniaxial compression is along the direction of nematic order, we find that the soft elasticity characteristic of liquid crystal elastomers leads to a  critical stress for wrinkling which} is very small compared to the case of an isotropic substrate. \trpa{We also determine the wavelength of the wrinkles at the critical stress, and show how the critical stress and wavelength depend on substrate depth and the anisotropy of the polymer chains in the nematic elastomer.}
 }
\end{abstract}



\maketitle
\section{Introduction}

\trpa{Wrinkles are easily created by compressing a thin film that coats an elastic substrate that is more compliant than the film~\cite{Li_etal2012}.  Since wrinkle patterns can have micron-scale periodicities, wrinkling is a promising phenomenon to exploit for many technological applications, such as diffraction gratings~\cite{jcprey2015}, dynamically tunable optical devices~\cite{aomAizenberg2013},  and smart adhesives~\cite{Chan_etal2008}.  A requirement for such applications is the ability to create ordered wrinkle patterns in a specified orientation. Some methods for creating wrinkles lead to disordered patterns. For example, wrinkles may be created by depositing a metal film on an elastomer substrate~\cite{Bowden_etal1998}. The heat of the deposition process expands the substrate and the film; upon cooling, the different thermal expansivities of the two materials leads to a biaxial compression of the film. The resulting stress is released by a wrinkle pattern in which the crests and troughs of the wrinkles form a disordered herringbone pattern~\cite{Bowden_etal1998}. Control of the orientation of the wrinkle pattern can be achieved by manipulating the topography; for example imposing a relief structure in the substrate such as an array of raised rectangles can lead to patches of one-dimensional wrinkle patterns, or, in the case of  an array of steps in the substrate, a one-dimensional wrinkle pattern throughout the whole film~\cite{Bowden_etal1998}.  }

\trpa{An alternate technique for controlling the wrinkle pattern is to use an anisotropic film or substrate.  For example, single crystal SiGe films buckled on a melted glass layer tend to have undulations parallel to the two orthogonal $\langle 100\rangle$ directions of the crystal film~\cite{Peterson_etal2006,Yu_etal2005}. The theory for wrinkling patterns of such an anisotropic crystal film on a viscoelastic substrate was studied in~\cite{ImHuang2008}. Another way to use anisotropy to create wrinkles with a desired orientation is to use a substrate of aligned liquid crystal polymer and cross-link the top layer of the substrate by exposing it to a plasma environment; the resulting wrinkles that form to release the stress in the cross-linked layer form parallel to the direction of the unpolymerized liquid crystal polymers in the substrate~\cite{Kang_etal2012}.  Finally, Agrawal and collaborators used a thin film atop a liquid crystal elastomer in the ordered state to create a one-dimensional pattern of wrinkles either perpendicular or parallel to the direction of order ~\cite{aditya2012SoftMatter,aditya2014SoftMatter}. The molecular order of the nematic causes an anisotropic distribution of the segments comprising the polymer chains of the elastomer, with the distribution elongated along the direction of alignment of the nematic order. Increasing the temperature reduces the nematic order, leading to a compressive stress in the film such that the wrinkles form perpendicular to the direction of molecular alignment~\cite{aditya2012SoftMatter}. When the temperature is decreases, nematic order increases, leading to a compressive stress perpendicular to the alignment direction via the Poisson effect and thus wrinkles with crests parallel to the direction of order.}

\trpa{In this paper, we develop the theory for small-amplitude wrinkles in an isotropic film on a nematic elastomer substrate, with the goal of calculating the critical compressive stress required to induce wrinkling and the wavelength of wrinkles at their onset. In section II, we review the von Karman equations for a thin film, and then briefly review the key results for small-amplitude wrinkling on an isotropic substrate. Then we introduce the elastic energy for a nematic elastomer substrate, and use this energy to calculate the critical stress and initial wrinkle wavelength in section III.  Since ideal nematic elastomers exhibit \textit{soft elasticity}, or nonzero deformation without energy cost~\cite{WarnerTerentjev},  we find that the critical stress for wrinkling is much smaller than the case of an isotropic elastomer substrate.  In section IV,  we discuss these results and compare them}  with the experimental findings. \trpa{I}n the last section we conclude our analysis and discuss 
future directions.

\section{Model}
\begin{figure}[t]
\includegraphics[width=
3in]{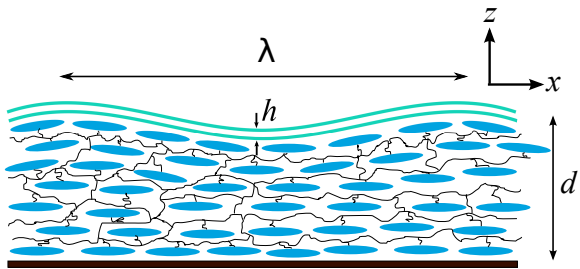}
\caption{(Color online.) Schematic diagram of a thin film of thickness $h$ deposited on a liquid crystal elastomer
 substrate of thickness $d$. The mesogens are not drawn to scale. The film has wrinkles of wavelength $\lambda=2\pi/q$, with the modulation along the $x$ direction.}
\label{setupfig}
\end{figure}

Fig. \ref{setupfig} shows a schematic diagram of a  thin film deposited on a liquid crystal elastomer substrate. Initially the film is flat, lying in the $xy$-plane, with the nematic elastomer beneath it in the half-space $z<0$, and with the directors aligned along the $x$-direction.  The thicknesses of the film and the nematic elastomer substrate are  $h$ and $d$, respectively. The film is subject to an initial in-plane compressive strain, and we \trpa{focus on} a one-dimensional sinusoidal  pattern of wrinkles with wavevector of magnitude $q$ along the $x$-direction. \trpa{Below we argue that wrinkles with a wavevector along the $y$ direction behave as wrinkles on an isotropic substrate.} First we review the von Karman equations for a thin film subject to an external normal force per unit area~\cite{landau_lifshitz_elas,HuangHongSuo2005}. \trpa{After reviewing the case} of a  thin film on an isotropic substrate, we derive the equations \trp{for force and torque balance for} the nematic elastomer substrate using both the ideal rubber and semi-soft models of nematic elastomers~\cite{WarnerTerentjev}.

\subsection{Equilibrium conditions for the thin film}

The thin film is modeled as an isotropic elastic plate with Young's modulus $E_\mathrm{f}$ and Poisson ratio $\nu_\mathrm{f}$. Hooke's law  is
\begin{equation}
	\varepsilon^\mathrm{f}_{ij}=\left[(1+\nu_\mathrm{f})\sigma^\mathrm{f}_{ij}-\nu_\mathrm{f}\sigma^\mathrm{f}_{ll}\delta_{ij}\right]/E_\mathrm{f},
\end{equation}
or
\begin{equation}
	\sigma^\mathrm{f}_{ij}=\frac{E_\mathrm{f}}{1+\nu_\mathrm{f}}\left[\varepsilon^\mathrm{f}_{ij}+\frac{\nu_\mathrm{f}}{1-2\nu_\mathrm{f}}\varepsilon^\mathrm{f}_{ll}\delta_{ij}\right],
\end{equation}
where $\sigma^\mathrm{f}_{ij}$ and $\varepsilon^\mathrm{f}_{ij}$ are the strain and stress tensors, respectively, and $i$ and $j$ run
over $x$, $y$, and $z$. \trp{We use the Einstein summation convention of summing over repeated indices.}
\trp{We impose the conditions of plane stress because we assume that in-plane stresses $\sigma^\mathrm{f}_{\alpha\beta}$ ($\alpha$, $\beta=x$, $y$) are much larger than the stresses $\sigma^\mathrm{f}_{iz}$ applied at the film-substrate interface.} The conditions of plane stress~\cite{landau_lifshitz_elas} $\sigma^\mathrm{f}_{iz}=0$ lead to 
\begin{equation}
	\sigma^\mathrm{f}_{\alpha\beta}=\bar{E}_\mathrm{f}\left[(1-\nu_\mathrm{f})\varepsilon^\mathrm{f}_{\alpha\beta}+\nu_\mathrm{f}\varepsilon^\mathrm{f}_{\gamma\gamma}\delta_{\alpha\beta}\right],
\end{equation}
where $\bar{E}_\mathrm{f}=E_\mathrm{f}/(1-\nu_\mathrm{f}^2)$ and $\alpha$ and $\beta$ run over $x$ and $y$.
Denoting the initial strain in the film as $\varepsilon^{\mathrm{f}0}_{\alpha\beta}$, the final \trpa{in-plane} strain is 
given 
\trpa{by the leading order symmetric gradients of the} in-plane displacements $u^\trpa{\mathrm{f}}_\alpha$ and out-of-plane displacement $u^\trpa{\mathrm{f}}_z$ by
\begin{equation}
	\varepsilon^\mathrm{f}_{\alpha\beta}=\varepsilon^{\mathrm{f}0}_{\alpha\beta}+\frac{1}{2}\left(\frac{\partial u^\mathrm{f}_\alpha}{\partial x_\beta}+\frac{\partial u^\mathrm{f}_\beta}{\partial x_\alpha}\right)+\frac{1}{2}\frac{\partial u^\mathrm{f}_z}{\partial x_\alpha}\frac{\partial u^\mathrm{f}_z}{\partial x_\beta}.
\end{equation}
The von Karman equations governing the equilibrium behavior of the film \trpa{follow from demanding that the sum of the bending energy, stretching energy, and work due to external force is stationary under variations in  the displacements~\cite{landau_lifshitz_elas}:} 
\begin{eqnarray}
	\label{vonkarman}
	D\nabla^4 u^\mathrm{f}_z-h\frac{\partial}{\partial x_\beta}\left(\sigma^\mathrm{f}_{\alpha\beta}\frac{\partial u^\mathrm{f}_z}{\partial x_\alpha}\right)\hsn{-\sigma_n}&=&0\label{fvk1}\\
	h\frac{\partial\sigma^\mathrm{f}_{\alpha\beta}}{\partial x_\beta}-T_{\alpha}&=&0,\label{fvk2}
\end{eqnarray}
where $D=\bar{E}_\mathrm{f} h^3/12$ is the bending modulus of the film, \hsn{$\sigma_n$} is the external  normal force per unit area exerted on the film\trpa{, arising from the deformed substrate}, and $T_\alpha$ is the $\alpha$ component of the tangential force per unit area exerted on the film. 
\trpa{Note that an implicit assumption of this approach is that the film thickness $h$ is much smaller than all other lengths in the problem, such as the wrinkle wavelength $2\pi/q$ and the substrate depth $d$. In the experiments that motivate us~\cite{aditya2014SoftMatter}, $h$ ranges from $30$\,nm to $400$\,nm, the substrate thickness $d=360\,\mu$m, and the wrinkle wavelength is observed to be range from 2\,$\mu$m to $40$\,$\mu$m as the thickness increases. Also, in these experiments, the Young's modulus of polystyrene is $E_\mathrm{f}=3.5$\,GPa, and the Young's modulus of the liquid-crystal elastomer is $E=0.24$\,MPa~\cite{aditya2014SoftMatter}.}

\subsection{\trpa{Scaling arguments for the critical stress and wrinkling wavenumber}}
\label{sec:scaling}
\trpa{Before turning to the nematic elastomer substrate, we review simple physical arguments for the critical compressive stress and wrinkle wavelength for the case of an isotropic substrate~\cite{biot1937,Herringbone2004}, using the scaling arguments of~\cite{prlMahadevan2013}. 
The substrate has an elastic energy ${\cal F}_\mathrm{s}=(k/2)\int\mathrm{d}A (u_z^\mathrm{f})^2$, leading to a normal stress $\sigma_\mathrm{n}=-ku_z^\mathrm{f}$, where $k$ is the effective stiffness of the substrate. For example, $k=\rho g$ for a liquid substrate, where $\rho$ is the liquid density and $g$ is the acceleration of gravity, and $k=2\mu q$ for an isotropic elastic substrate, where $\mu$ is the shear modulus of the incompressible substrate. 
We look for wrinkling solutions $u_z^\mathrm{f}=A\cos qx$ to Eq.~(\ref{fvk2}), which takes the form $\Delta(q) A=0$, with 
\begin{equation}
\Delta(q)=\left(D q^4 +h \sigma_{xx}^\mathrm{f} q^2+k\right).
\end{equation}
To get a nonzero wrinkling amplitude $A$ requires $\Delta=0$, or, equivalently, a compressive film stress along $x$ of
\begin{equation}
\sigma_0=-\frac{D q^4+k}{h q^2}.\label{compstress}
\end{equation}
The nature of the  critical compressive stress $\sigma_\mathrm{c}$, which is the minimum value of $\sigma_0$, or the value at the onset of wrinkling,  and the corresponding wavenumber $q_\mathrm{c}$  is determined by how the stiffness $k$ depends on $q$ for small $q$.  If $k\propto q^n$ with $0\le n<2$, then we can find $q_\mathrm{c}$ by balancing the 
bending forces with the substrate forces, to find~\cite{prlMahadevan2013}
\begin{equation}
q_\mathrm{c}\sim\left(\frac{k}{D}\right)^{1/4}.
\end{equation}
Since the compressive stress in the film also leads to forces that must balance with the bending forces, we  have $h\sigma_\mathrm{c} q_\mathrm{c}^2\sim D q_\mathrm{c}^4$, or a critical compressive film stress with magnitude~\cite{prlMahadevan2013}
\begin{equation}
|\sigma_{\mathrm{c}}|\sim\frac{1}{h}\sqrt{k D}.
\end{equation}
For example, in the case of an incompressible isotropic rubber substrate, the elastic free energy density is $\mu \varepsilon_{ij}\varepsilon_{ij}$, where $\varepsilon_{ij}$ is the strain tensor of the substrate, $\mu=E/3$ is the shear modulus for the substrate, and $E$ is the Young's modulus of the substrate~\cite{landau_lifshitz_elas}.
Assuming that the penetration depth for the deformation in the wrinkled substrate is $1/q$ leads to $k\sim Eq$ and
\begin{equation}
q_\mathrm{c}h\sim\left(\frac{E}{E_\mathrm{f}}\right)^{1/3},\label{qscaling}
\end{equation}
and a critical compressive film stress with magnitude
\begin{equation}
|\sigma_{\mathrm{c}}|\sim E_\mathrm{f}\left(\frac{E}{E_\mathrm{f}}\right)^{2/3}.\label{sigmascaling}
\end{equation}
The critical wavelength is proportional to the film thickness, since in the limit of an infinitely deep substrate there are no other material length scales in the problem. Also, the critical stiffness increases when either the film stiffness or the substrate stiffness increases, but it is independent of the film thickness since again there is no other material length scale.}

\trpa{For a thin substrate with $qd\ll1$, the constraint of incompressibility leads to a lateral deformation of the form $u_x\sim u_z/(qd)$, and thus a dominant strain of $\partial_z u_x\sim u_z/(qd^2)$. Balancing the substrate deformation energy with the film bending energy leads to 
\begin{equation}
q_\mathrm{c}h\sim\left({\frac{h}{d}}\right)^{1/2}\left(\frac{E}{E_\mathrm{f}}\right)^{1/6},
\end{equation}
with a corresponding critical compressive film stress
\begin{equation}
|\sigma_{\mathrm{c}}|\sim E_\mathrm{f}\frac{h}{d}\left(\frac{E}{E_\mathrm{f}}\right)^{1/3}.
\end{equation}
A more detailed analysis reveals that in the thin substrate limit, the critical stress and wavenumber depend sensitively on the Possion ratio of the substrate, and therefore more care must be taken when the substrate is thin and nearly incompressible~\cite{HuangHongSuo2005}.
}

\trpa{If the stiffness $k\propto q^n$ with $n\ge 2$ for small $q$, then we get different scaling laws. If $n=2$, as we shall see below occurs in the case of a nematic elastomer with the Frank energy included, the quadratic dependence of $\Delta(q)$ on $q$ at small $q$ implies that $q_\mathrm{c}=0$. The balance of stretching and substrate forces in this case leads to a compressive stress of magnitude
\begin{equation}
|\sigma_{\mathrm{c}}|\sim\frac{k}{h}\label{kisqsquared}
\end{equation}
for a thick substrate. 
If $n>2$, then $q_\mathrm{c}=0$ and $\sigma_{\mathrm{c}}^\mathrm{f}=0$. 
}


\subsection{Equilibrium conditions for the liquid crystal elastomer substrate}

We consider a nematic liquid crystal elastomer with an elastic deformation described by $\mathbf{X}=\mathbf{x}+\mathbf{u}$, from an \trpa{undeformed configuration} with material points \trpa{at locations} $\mathbf{x}$ to a \trpa{deformed configuration} with material points $\mathbf{X}$, and with a deformation field $\mathbf{u}$. The deformation tensor is given by $\lambda_{ij}=\partial X_i/\partial x_j$.  The director field 
\trpa{in the undeformed configuration} is given by  $\mathbf{n}_0$, and transforms to $\mathbf{n}$ upon the elastic deformation. \trp{Considering the polymer chains of the elastomer as random walks,} the nematic order leads to an anisotropic step length tensor $\ell_{ij}=\ell_{||}n_i n_j+\ell_\perp(\delta_{ij}-n_in_j)$, with a similar expression 
\trpa{$\ell_{0ij}$} for the step length tensor 
\trpa{in the undeformed configuration}\trp{~\cite{WarnerTerentjev}}\footnote{\trpa{Although the thermal and strain history of a given sample may be unknown, if we measure strains relative to the relaxed state, then we need not distinguish between the relaxed state and the state of formation~\cite{WarnerTerentjev}.}}.
The step length is the roughly the length over which the polymer chain can bend. Parallel and perpendicular refer to directions parallel and perpendicular to the director, respectively. We assume that the magnitude of the nematic order parameter is unchanged by the elastic deformation and thus \hsn{$\ell_{\trpa{0}||}=\ell_{||}$, and $\ell_{\trpa{0}\perp}=\ell_\perp$.}

For an incompressible nematic elastomer, classical rubber elasticity theory modified to account for the anisotropic step length tensor leads to the free energy density~\cite{WarnerTerentjev},
\begin{equation}
	F=\frac{\mu}{2}\ell_{0ij}\lambda_{kj}\ell^{-1}_{kl}\lambda_{li},
\end{equation}
where $\mu$ is the shear modulus.  We will study small deformations away from a \trpa{relaxed} equilibrium configuration with $\mathbf{u}=0$ and $\mathbf{n}=\mathbf{n}_0$. Expanding the free energy to second order in $u_{ij}$ and $\delta\mathbf{n}=\mathbf{n}-\mathbf{n}_0$, and using the constraint of incompressibility
\begin{equation}
	1=\mathrm{det}\lambda_{ij}=1+u_{ii}-\frac{1}{2}u_{ij}u_{ji}+\cdots,
\end{equation}
with $u_{ij}=\partial u_i/\partial x_j$, 
yields the free energy density to quadratic order in strain and director rotation~\cite{WarnerTerentjev} 
\begin{eqnarray}
	F&=&\mu\mathrm{Tr}(\boldsymbol{\varepsilon}\cdot\boldsymbol{\varepsilon})+\mu\frac{(r-1)^2}{2r}\left[\mathbf{n}\cdot\boldsymbol{\varepsilon}\cdot\boldsymbol{\varepsilon}\cdot\mathbf{n}-(\mathbf{n}\cdot\boldsymbol{\varepsilon}\cdot\mathbf{n})^2\right]\nonumber\\
	&+&\frac{D_1}{2}\left[\left(\boldsymbol{\Omega}-\boldsymbol{\omega}\right)\times\mathbf{n}\right]^2+D_2\mathbf{n}\cdot\boldsymbol{\varepsilon}\cdot(\boldsymbol{\Omega}-\boldsymbol{\omega})\times\mathbf{n}.\label{WTF}
\end{eqnarray}
In Eq.~(\ref{WTF}) we have introduced the linear strain tensor of the elastomer, $\varepsilon_{ij}=(u_{ij}+u_{ji})/2$, the step length anisotropy ratio $r=\ell_{||}/\ell_\perp$ ($r>1$ for prolate molecules), the material rotation $\Omega_i=(1/2)(\boldsymbol{\nabla}\times\mathbf{u})_i$, and the director rotation $\boldsymbol{\omega}$, which is related to the change in director via $\delta\mathbf{n}=\boldsymbol{\omega}\times\mathbf{n}_0$.   To this order of approximation, we make no errors writing $\mathbf{n}$ for $\mathbf{n}_0$ in the free energy. 

The elastic moduli $D_1$ and $D_2$ are given in the ideal rubber theory by $D_1=\mu(r-1)^2/r$ and  \trp{$D_2=\mu(r^2-1)/r$}~\cite{WarnerTerentjev}. 
If we take into account  semi-soft elasticity,  the modulus $D_1$ will be modified and  there can be a contribution  to $D_1$ proportional to $r-1$ ~\cite{WarnerTerentjev}. For example, fluctuations in the effective order $Q_\mathrm{eff}=(1+\delta)Q$ experienced by a polymer chain, arising from compositional fluctuations in the polymers that make up the elastomer, lead to a modulus~\cite{WarnerTerentjev}
\begin{equation}\label{d1}
	D_1=\trp{\mu\dfrac{\det{({\ell}_{0})_{ij}}}{\det{{\ell}_{ij}}}\left[\dfrac{(r-1)^2}{r}+a_1(r-1)\right]},
\end{equation}
where 
\begin{equation}
	a_1=3\dfrac{\left\langle\delta^2\right\rangle}{r+2}     \dfrac{Q_0(3+2Q_0)}{(1+2Q_0)(1-Q_0)}.
\end{equation}
\trpa{Again} the subscript \hsn{$0$} denotes the values when the elastomer is formed, $Q=(r-1)/(r+2)$ is the nematic order parameter of the nematic rubber substrate\trpa{,} and $\left\langle \delta^2 \right\rangle$ represents the compositional-induced fluctuations in $Q$. \hsn{We assume that $\det{({\ell}_{0})_{ij}}/\det{({\ell})_{ij}} = 1$, which is a good approximation for the small thermal changes.} 
In the limit $|r-1|\gg r a_1$, the second term in Eq. \eqref{d1} can be neglected and the ideal rubber theory is sufficient to explain the elastic properties of the nematic rubber substrate. We consider this limit in the present section and discuss the role of semi-softness in Sec.~\ref{smallr11}.

The stress tensor in the elastomer is given by~\cite{landau_lifshitz_elas} 
\begin{equation}
	\sigma_{ij}=\frac{\partial F}{\partial u_{ij}}.\label{stress}
\end{equation}
Given the anisotropy induced by the director, we must be careful about the order of indices; the condition for equilibrium  compatible with the convention of Eq. \eqref{stress}  is $\partial\sigma_{ik}/\partial x_k=0$. Carrying out the differentiation in Eq. \eqref{stress} leads to a stress of the from $\sigma_{ij}=-p\delta_{ij}+\sigma^\mathrm{d}_{ij}$, with
\begin{eqnarray}
	\sigma^{\mathrm{d}}_{ij}/\mu&=&2\varepsilon_{ij}+\left(\frac{1}{r}-1\right)n_i\varepsilon_{jl}n_l+(r-1)n_j\varepsilon_{il}n_l\nonumber\\ &-&\frac{(r-1)^2}{r}\mathbf{n}\cdot\boldsymbol{\varepsilon}\cdot\mathbf{n}\,n_in_j+{\frac{(r-1)^2}{2r}\epsilon_{kji}(\Omega-\omega)_k}\nonumber\\
	&+&\frac{r^2-1}{2r}\left\{n_i\left[\left(\boldsymbol{\Omega}-\boldsymbol{\omega}\right)\times\mathbf{n}\right]_j\right.\nonumber\\
	&&\qquad \qquad \qquad+\big. n_j\left[\left(\boldsymbol{\Omega}-\boldsymbol{\omega}\right)\times\mathbf{n}\right]_i \Big\},\label{fullstress}
\end{eqnarray}
where $p$ is the pressure field enforcing incompressibility ($\varepsilon_{ii}=0$ at this order) and $\epsilon_{ijk}$ is the three-dimensional antisymmetric symbol.  Note that in deriving Eq. \eqref{fullstress}, we have assumed that $\boldsymbol{\Omega}$ and $\boldsymbol{\omega}$ are perpendicular to the plane in which gradients in $\mathbf{u}$ and $\delta\mathbf{n}$ lie.

Now we consider a semi-infinite liquid crystal elastomer, with the free surface lying in the $xy$ plane; the elastomer lies in the region $z<0$. When the elastomer and deposited film are cooled within the nematic phase, wrinkles appear with their ridges and troughs parallel to $\mathbf{n}_0$~\cite{aditya2012SoftMatter}, and the system is equivalent to the isotopic case, $r=1$. The more interesting case occurs when the system is heated within the nematic phase. As the temperature increases, $Q$ decreases and the nematic elastomer changes shape, contracting along $\mathbf{n}_0$ and  producing wrinkles with ridges oriented perpendicular to the director (Fig.~\ref{setupfig})~\cite{aditya2012SoftMatter}.   Assuming $\mathbf{n}_0=\hat{\mathbf{x}}$ and translational invariance along 
\trpa{the} $y$ direction, the stress tensor simplifies drastically:
\begin{eqnarray}
	\sigma_{xx}&=&-p+2\mu u_{xx}\\
	\sigma_{zz}&=&-p+2\mu u_{zz}\label{zstress}\\
	\sigma_{xz}&=&\mu\left[\frac{1}{r}u_{xz}+u_{zx}+\frac{1-r}{r}\theta\right]\\
	\sigma_{zx}&=&r \sigma_{xz}=\mu\left[r u_{zx}+u_{xz}+(1-r)\theta\right],
\end{eqnarray}
where $\theta$ is the tilt angle of the director in the $x$-$z$ plane, measured from the $x$ direction: \trp{$\mathbf{n}=\mathbf{n}_0+\delta\mathbf{n}=\hat{\mathbf x}+\theta\hat{\mathbf z}$}.

The stress tensor is not symmetric because the director field can transmit torques across surfaces. Therefore, in addition to the force balance equation $\partial\sigma_{ik}/\partial x_k=0$, we must enforce torque balance, \trp{$\delta \mathcal{F}/\delta\theta=0$, where $\mathcal{F}$ is the total energy}, including the  Frank nematic elastic energy~\cite{deGennesProst}. \trp{While Frank elasticity is often not considered in nematic elastomers since its effects are dominated by the network elasticity and nematic-network coupling~\cite{WarnerTerentjev}, we will see below that Frank elastic plays an important role in determining the size of the region of the elastomer that gets deformed.}  Within the single elastic constant approximation the Frank energy density is given by
\begin{equation}
	F_\mathrm{F}=\frac{K}{2}\frac{\partial\theta}{\partial x_i}\frac{\partial\theta}{\partial x_i}.
\end{equation}
This energy density generates a contribution to the stress tensor depending on the configuration of the directors~\cite{deGennesProst,landau_lifshitz_elas}. Using the Ericksen formulation \cite{deGennesProst} for this contribution $\sigma^\mathrm{F}_{ik}$ to the stress tensor, we vary $\mathbf{u}$ and $\mathbf{n}$ independently and obtain
\begin{equation}
	\sigma^\mathrm{F}_{ik}=-K\frac{\partial\theta}{\partial x_i}\frac{\partial\theta}{\partial x_k}.\label{Fstress}
\end{equation}
To linear order in $\theta$ the Ericksen stress does not contribute to the equations for force balance.
Torque balance takes the form
\begin{equation}\label{torquebal}
	-\frac{K}{\mu}\nabla^2\theta+\frac{1-r}{r}\left[(1-r)\theta+r u_{zx}+u_{xz}\right]=0.
\end{equation}
 \trpa{Note how the torque from the directors leads to an antisymmetric part of the stress tensor: $-K\nabla^2\theta+(1-r)(\sigma_{xz}-\sigma_{zx})=0$.}
Note \trpa{also} that the Frank constant introduces a new length scale, $l=\sqrt{K/\mu}$. 
Using typical values of $\mu\approx10^5$\,Pa and $K\approx10^{-11}$\,N, we find that $l\approx10$\,nm.  Since the Frank constant is typically of the order of the square of the order parameter $Q$~\cite{deGennesProst}, we take
\begin{equation}
	K=\kappa\left(\frac{r-1}{r+2}\right)^2,
\end{equation}
where $\kappa\approx10^{-11}$\,N.  It is convenient to define an $r$-independent length scale $l_0=\sqrt{\kappa/\mu}$; this length scale is approximately $l_0\approx100\,$nm.

\section{Results}\label{results}
In this section we solve the force and torque-balance equations for the elastomer substrate to determine the normal force per area $\sigma_n$ acting on the wrinkled film. We treat several cases: an ideal nematic elastomer substrate with Frank elasticity disregarded, an ideal nematic elastomer substrate with Frank elasticity included, and finally the limit of vanishing anisotropy in which semi-soft elasticity becomes prominent. In all cases, once the normal force is found, we use the von Karman equations (\ref{fvk1}--\ref{fvk2}) for the thin film to determine the critical compressive film stress at which wrinkling occurs, as well as the wrinkle wavelength.


\trpa{\subsection{Ideal nematic elastomer substrate without Frank elasticity}\label{nofrank}
To highlight the role of soft elasticity in the problem of wrinkling on a nematic liquid crystal elastomer, we consider the simplest case in which the nematic elastomer is ideal, and Frank elasticity is disregarded. If the elastomer is of infinite depth, then the stiffness of the substrate vanishes, $k=0$. For a qualitative explanation, recall that the polymer chains of the elastomer have an ellipsoidal distribution with major axis along the local director. Since there is no entropy cost for rotating the ellipsoids, there is no free energy cost for rotating the ellipsoids to make the elastomer accommodate the wrinkle deformation of the thin film. Thus, the soft elasticity of the substrate implies that the ideal nematic elastomer effectively behaves like a liquid substrate~\cite{Huang_etal2007} in the small-amplitude wrinkling problem. }

\trpa{We can examine this problem more quantitatively by considering the case of finite depth $d$. When $K=0$, the torque balance Eq.~(\ref{torquebal}) implies that the shear stresses vanish, $\sigma_{zx}=r\sigma_{xz}=0$. Thus, the force balance equations $\partial_k\sigma_{ik}=0$ simplify considerably. With the boundary conditions $u_z(z=0)=A\cos qx$ and $u_z(z=-d)=0$, we find
\begin{eqnarray}
u_x&=&-\frac{A}{qd}\sin(qx)\label{idealux}\\
u_z&=&A\left(1+\frac{z}{d}\right)\cos(qx)\label{idealuz}\\
\theta&=&\frac{r}{1-r}qA\left(1+\frac{z}{d}\right)\sin(qx)\label{idealtheta}\\
\sigma_{zz}&=&\frac{4\mu A}{d}\cos(q x).\label{idealsigzz}
\end{eqnarray}
Note that since the shear stress vanishes in the bulk in this idealized case, the boundary conditions at $z=0$ and $z=d$ must be $\sigma_{xz}=0$ instead of a condition on $u_x$.} \trpc{Also, Eqn.~(\ref{idealtheta}) reveals that our small-amplitude expansion breaks down when $r\rightarrow1$.}

\trpa{The stiffness $k$ can be read off from Eq.~(\ref{idealsigzz}), or equivalently, by computing the free energy (per unit length in the $y$ direction) from the expressions in Eqs.~(\ref{idealux}--\ref{idealtheta}), which gives $\mathcal{F}_\mathrm{S}=2\mu A^2\pi/(qd)$. Therefore, the stiffness is $k\sim E/d$, leading to a critical wavelength and critical compressive stress of
\begin{eqnarray}
q_\mathrm{c}h&\sim&\left(\frac{h}{d}\right)^{1/4}\left(\frac{E}{E_\mathrm{f}}\right)^{1/4}\label{qcsmalld}\\
|\sigma_{\mathrm{c}}|&\sim&\left({\frac{h}{d}}\right)^{1/2}E_\mathrm{f}\left(\frac{E}{E_\mathrm{f}}\right)^{1/2}.\label{sigmacsmalld}
\end{eqnarray}
The critical stress in this idealized model can be made much smaller than the stress in the isotropic case by making the depth much greater than the film thickness, $d\gg h$. In this limit the critical stress vanishes, as argued above.
}

\trpa{Although this idealized model correctly captures the fact that the soft elasticity of a nematic elastomer leads to a greatly reduced critical stress for wrinkling, it also has some shortcomings. For example, it predicts that for an infinitely thick substrate, the deformation penetrates infinitely far into the bulk, since the deformation and angle field become independent of $z$ despite being non-vanishing in this limit. In other words the penetration depth, denoted by $l_\mathrm{p}$, is infinite. However, the resulting bend deformation of the director field would have an infinite energy cost for any nonzero Frank constant, no matter how small. Thus, we expect the Frank constant to lead to a finite penetration depth for the deformation and director distortion. We consider the effects of the Frank constant in the next subsection.  Also, the \trpc{displacements and the stress in} (\ref{idealux}), (\ref{idealuz}), and (\ref{idealsigzz}) do not have the correct isotropic limits when $r\to1$, \trpc{and as already pointed out, the angle field $\theta$ diverges when $r\rightarrow1$}. This shortcoming ultimately arises from our assumption of idealized soft elasticity, and will be rectified when we discuss semi-soft elasticity below.  }

\subsection{Ideal nematic elastomer substrate with Frank elasticity}

\trpa{Now we consider the governing equations with nonzero Frank constant.}
\trpa{Since the substrate is taken to be incompressible, we introduce the stream function $\psi$ such that $\boldmath{\nabla}\times(-\psi\hat{\mathbf{y}})=\mathbf{u}$.}
Assuming that the solution has the 
form
\begin{eqnarray}
	\psi(x,z)&=&\Psi(z)\sin(qx)\\
	\theta(x,z)&=&\Theta(z)\sin(qx),
\end{eqnarray}
\trpa{the curl of the} force balance equations, \trpa{$\epsilon_{ijk}\partial_j\partial_l\sigma_{kl}=0$}, and torque balance equation, Eq. \eqref{torquebal}, take the form
\begin{eqnarray}
	\label{char_eq1}
	r q^4\Psi-2q^2\Psi''+\frac{1}{r}\Psi''''\qquad \qquad \quad \qquad \qquad &&\nonumber\\+(1-r)\left(\frac{1}{r}\Theta''+q^2\Theta\right)&=&0\\
	\label{char_eq2}
	(1-r)\left(\frac{1}{r}\Psi''+q^2\Psi\right)-l^2(\Theta''-q^2\Theta)\qquad   &&\nonumber\\+\frac{(1-r)^2}{r}\Theta&=&0,
\end{eqnarray}
where the primes denote differentiation with respect to $z$. 
Note that $l=l_0(r-1)/(r+2)$, and $ql_0\ll1$.

Assuming $\Psi$ and $\Theta$ are proportional to $\exp(\alpha z)$ leads to the characteristic equation \trpa{for $\alpha$},
\begin{equation}
	\left|
	\begin{array}{cc}
		{(\alpha^2-r q^2)^2}/{r} &(1-r)\left(q^2+{\alpha^2}/{r}\right)     \\
		(1-r)\left(q^2+{\alpha^2}/{r} \right)  &  \trpa{(1-r)^2\left[\frac{1}{r}-\frac{l_0^2(\alpha^2-q^2)}{(r+2)^2}\right]} 
	\end{array}
	\right|=0.\label{detform}
\end{equation}
Excluding the solution $r=1$ of Eq.~\eqref{detform}\trpa{, which corresponds to the isotropic case,} the characteristic equation reduces to
\begin{eqnarray}
	\alpha^6-(1+2r)q^2\alpha^4+\left[r q^2+4\frac{\hsn{(r+2)}}{l_0^2}\right](r+2)q^2\alpha^2\nonumber\\-r^2q^6=0.\qquad\label{charac}
\end{eqnarray}
 \trpa{The six roots of (\ref{charac}) are
 \begin{equation}
	\alpha=\left\{\begin{array}{c}\pm 1/l_\mathrm{p}\\ \pm(1\pm\mathrm{i})/\delta\end{array}\right. ,\label{approxroots}
\end{equation}
where}
\begin{eqnarray}
1/l_\mathrm{p}&\approx& r q^2 l_0/[2(r+2)]\label{eqn:lp}\\
1/\delta&\approx&q\sqrt{(r+2)/({ql_0})}
\end{eqnarray}
\trpa{for $ql_0\ll1$. The first two roots correspond to a very large penetration depth $l_\mathrm{p}$, much longer than the penetration depth in the case of an isotropic substrate: $\l_\mathrm{p}\sim1/(q^2l_0)\gg1/q$. The remaining roots correspond to a thin boundary layer of thickness $\delta\sim\sqrt{l_0/q}\ll 1/q$ near the top and bottom interfaces of the substrate, where the director field and deformation adjust from the bulk values to the values prescribed by the boundary conditions. }

\trpa{We can obtain the length scale $l_\mathrm{p}$ from a simple scaling argument. The soft elasticity of an ideal nematic liquid crystal elastomer implies that the shear stresses $\sigma_{zx}$  and $\sigma_{xz}$ vanish in our problem when the Frank constant is set to zero. For a nonzero Frank constant, the torque balance equation implies that $\sigma_{zx}\sim\mu q^2 l_0^2\theta\ll1$. On the other hand, the individual terms in the torque balance equation such as $u_{xz}$, $u_{zx}$, and $\theta$, are not all small; therefore the dominant terms of the shear stress must balance each other. Assuming we consider a region of the elastomer far from the film, where the boundary layer terms have no effect, the rate of variation in the $x$ direction is $q$, and the rate of variation in the $z$ direction is $1/l_\mathrm{p}$. Balancing the dominant terms $u_{zx}$ and $\theta$ of the shear stress yields $\theta\sim q u_z$. (To see that the term $u_{xz}$ is subdominant, use incompressibility $q u_x\sim u_z/l_\mathrm{p}$ and $ql_\mathrm{p}\gg1$.) On the other hand, the balance of $z$ forces, $\partial_x\sigma_{zx}+\partial_z\sigma_{zz}=0$, implies $q\sigma_{zx}\sim \sigma_{zz}/l_\mathrm{p}$, or
\begin{equation}
q (l_0^2 q^2\theta)\sim\frac{1}{l_\mathrm{p}^2}u_z.
\end{equation}
Therefore, the penetration depth is given by 
\begin{equation}
\frac{1}{l_\mathrm{p}}\sim q^2 l_0.
\end{equation}}

\trpa{\subsubsection{Limit of a very thick substrate}}
\trpa{To clearly identify how the different length scales $1/q$, $l_\mathrm{p}$, and $\delta$ enter the solution, we first consider the case of a very thick substrate, $q d\gg 1$. In this case the boundary conditions are
\begin{eqnarray}
u_z(z=0)&=&A\cos(qx)\label{topbc1}\\
\theta(z=0)&=&-Aq\sin(qx)\label{topbc2}\\
\sigma_{xz}(z=0)&=&0,\label{topbc3}
\end{eqnarray}
with $u_\alpha$ and $\theta$ vanishing at $z\rightarrow-\infty$. 
\trpa{We assume zero} shear stress at the interface between the nematic elastomer substrate and the film. \trp{The approximation of zero shear stress at the film-substrate interface is common in studies of wrinkling~\cite{Herringbone2004}.}
Using these boundary conditions to determine $\Psi$ and $\Theta$, and then expanding the resulting coefficients to leading order in $q l_0$ leads to  
\begin{eqnarray}
\psi&\sim&\frac{A}{q}\sin(q x)\left[-{\mathrm{e}^{z/l_\mathrm{p}}}+\frac{1}{2}{q^2\delta^2}{\mathrm{e}^{z/\delta}}\sin(z/\delta)\right]\label{psiasympt}\\
\theta&\sim&\frac{Aq\sin(q x)}{1-r}\left[r\mathrm{e}^{z/l_\mathrm{p}}-\mathrm{e}^{z/\delta}\cos(z/\delta) \right].\label{thetaasympt}
\end{eqnarray}}
\trpa{Note the form of the stream function $\psi$. The second term of (\ref{psiasympt}) has a small amplitude and rapid decay compared to the the first term. However, we cannot neglect it relative to the first term when computing the displacements and strains, since the rapid decay length offsets the small amplitude of the second term. In fact it is safest to calculate the displacements, strains, and angle field first, and then expand the coefficients to leading order in $ql_0$. Doing so and using the force balance equation $\partial_i\sigma_{zi}=0$ to calculate $\sigma_{zz}$ leads to
\begin{equation}
\sigma_{zz}(z=0)=4\mu u_z^\mathrm{f}/l_\mathrm{p}+\mathcal{O}[(q l_0)^{3/2}].
\end{equation} }
Thus, for a thick substrate, $d\gg1/(q^2l_0)$, the stiffness of the elastomer is
\begin{equation}
k\sim \mu/l_\mathrm{p}\sim E/l_\mathrm{p},
\end{equation}
which is much softer than the isotropic stiffness, $k\sim Eq$.  As discussed in Sec.~\ref{sec:scaling}, the $q^2$ dependence of $k$ through Eq.~(\ref{eqn:lp}) implies that $q_\mathrm{c}=0$. (Ultimately, the finite size of the film in the $x$ direction will determine the critical wavenumber in this case.) Using the scalings of Eq.~(\ref{kisqsquared}) yields
\begin{equation}
|\sigma_{\mathrm{c}}|\sim E\frac{l_0}{h}\sim\frac{\sqrt{EK}}{h},
\end{equation}
which is independent of the properties of the film. \trpc{Alternately, since $k\propto q^2$ at small $q$, we can get the critical stress directly from Eq.~(\ref{compstress}):}
\begin{equation}
\trpc{-\sigma_{\mathrm{c}}h=\frac{2r}{r+2}\sqrt{\kappa E/3}.}\label{sigmaxxc}
\end{equation}
Although it may seem surprising that the critical stress diverges as $h\to0$, note that the critical compressive force per unit length remains finite.

\trpa{\subsubsection{Limit of a very thin substrate}

When the substrate is thin,  $qd\ll1$, then the penetration depth $l_\mathrm{p}\gg d$, since $d/l_\mathrm{p}\sim q^2 l_0 d\ll 1$. When the penetration depth is much larger than the size of the substrate, we can disregard the Frank energy. Thus, the limit $qd\ll1$ is exactly the same as the case of an nematic elastomer in the absence of Frank energy, considered in Sec.~\ref{nofrank}.}

\trpa{\subsubsection{Substrate of arbitrary thickness}}

\trpa{Since} the decay length $l_\mathrm{p}$ 
can \trpc{in principle} be of order the thickness $d$ of the substrate \trpa{when $ql_0\ll1$, we consider a substrate of finite thickness.}
\trpa{Also  we suppose that the director is parallel to the film-substrate interface and to  the flat base of the substrate, so that in addition to the conditions (\ref{topbc1}--\ref{topbc3}) , we have}
\begin{eqnarray}
\theta(z=-d)&=&0\\
u_x(z=-d)&=&0\\
u_z(z=-d)&=&0.\label{bound2}
\end{eqnarray}
\trpa{Calculating the displacements, angle field, strain, and stresses for small $q l_0$ (but not small $q^2 l_0 d\sim d/l_\mathrm{p}$) leads to 
\begin{equation}
\sigma_{zz}(z=0)=4\frac{\mu}{l_\mathrm{p}}u_z^\mathrm{f}\coth(d/l_\mathrm{p}).\label{fullstiff}
\end{equation}
}

\begin{figure}
\includegraphics[width=0.5\textwidth]{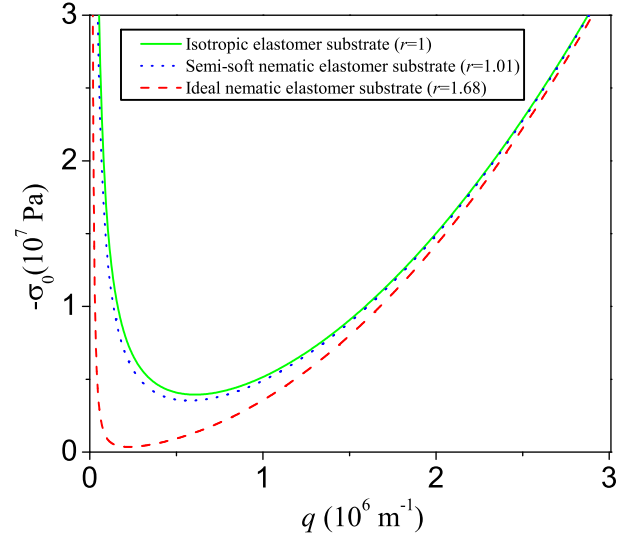}
\caption{\hsn{(\trpc{C}olor online).  Compressive stress $-\sigma_0$ as \trpc{a} function of \trpc{wrinkle} wavenumber  $q$ for \trpc{an ideal} nematic elastomer substrate with $r=1.68$ (red dashed line), \trpc{a} semi-soft nematic elastomer substrate with $r=1.01$ (blue dotted line), and \trpc{an} isotropic elastomer substrate with $r=1$ (green solid line), all with a shear modulus \trpc{$\mu=0.24\,$MPa} and \trpc{a polystyrene film with thickness $h=100$\,nm and stiffness $E_\mathrm{s}=3.5$\,GPa.}
For each curve, the critical stress $\sigma_\mathrm{c}$  corresponds to the minimum at $q_\mathrm{c}$; the critical stress for the nematic elastomer substrate is much smaller than for the isotropic elastomer substrate.
} }
\label{sigma_q}
\end{figure}

\trpa{Thus, wrinkles of wavenumber $q$ first form when the compression attains the value [recall Eq.~(\ref{compstress})],
\begin{equation}
-\sigma_{0}h=Dq^2+4\frac{\mu}{q^2l_\mathrm{p}}\coth(d/l_\mathrm{p}).
\end{equation}
Figure~\ref{sigma_q} shows the compressive stress $\sigma_0$ required to form wrinkles of wavenumber $q$ for a nematic elastomer substrate with $r=1.68$, and, for comparison, an isotropic elastomer substrate and a semi-soft nematic elastomer substrate \trpc{(discussed below in Sec.~\ref{smallr11})} with $r=1.01$. }

\begin{table*}
\begin{center}
\begin{tabular}{|l|c|c|}
\hline
 \multicolumn{1}{|c|}{\textrm{system}}  & $-\sigma_{\mathrm{c}}$ & $q_\mathrm{c}h$\\
\hline
\hline
\textrm{thick isotropic substrate, $qd\gg1$~\cite{Herringbone2004}} &$ ({1}/{3})E\left({4E}/{\bar{E}_\mathrm{f}}\right)^{1/3}$ & $\left({4 E}/{\bar{E}_\mathrm{f}}\right)^{1/3}$\\
\hline
\textrm{thin isotropic substrate, $qd\ll1$~\cite{HuangHongSuo2005}}& $\bar{E}_\mathrm{f}\left[{E}/{(3\bar{E}_\mathrm{f})}\right]^{1/3}{h}/{(4d)}$&$\left({24E}/{\bar{E}_\mathrm{f}}\right)^{1/6}\sqrt{{h}/{d}}$ \\
\hline
\textrm{thick nematic elastomer substrate without Frank energy} &$0$ &$0$ \\
\hline
\textrm{thin nematic elastomer substrate without Frank energy} &$(2/3)\sqrt{E\bar{E}_\mathrm{f}}\sqrt{h/d}$ &$2(E/\bar{E}_\mathrm{f})^{1/4}(h/d)^{1/4}$ \\
\hline
\textrm{thick nematic elastomer substrate with Frank energy $d/l_\mathrm{p} \gg1$} & $2r\sqrt{E\kappa/3}/[h(r+2)]$&$0$ \\
\hline
\textrm{thin nematic elastomer substrate with Frank energy $d/l_\mathrm{p} \ll1$} & $(2/3)\sqrt{E\bar{E}_\mathrm{f}}\sqrt{h/d}$&$2(E/\bar{E}_\mathrm{f})^{1/4}(h/d)^{1/4}$\\
\hline
\end{tabular}
\end{center}
\caption{\trpc{Summary of the critical compressive stress and critical wrinkling wavenumber for a thin film on an isotropic or nematic elastomer. In all cases the elastomer is incompressible and ideal. }}
\label{tablescales}
\end{table*}%
\trpa{The first wrinkles that form as the compression increases from zero is given by the critical wavenumber $q_c$ that minimizes $-\sigma_{0}(q)$, namely
\begin{equation}
\trpc{q_\mathrm{c}^2=\frac{2(2+r)}{r l_0 d}\sinh^{-1}\left(\frac{r}{r+2}\sqrt{\frac{12\kappa d}{\bar{E}_\mathrm{f}h^3}}\right),}\label{fullqc}
\end{equation}
with a corresponding critical stress of
\begin{eqnarray}
&-&\sigma_{\mathrm{c}}h=\frac{1}{3}\sqrt{\frac{E \bar{E}_\mathrm{f}h^3}{d}}\sqrt{1+\frac{r^2}{(r+2)^2}\frac{12\kappa d}{\bar{E}_\mathrm{f}h^3}}\nonumber\\
&+&\frac{\bar{E}_\mathrm{f}h^3}{6d}\sqrt{\frac{E}{3\kappa}}\frac{2+r}{r}\sinh^{-1}\left(\frac{r}{r+2}\sqrt{\frac{12\kappa d}{\bar{E}_\mathrm{f}h^3}}\right).
\end{eqnarray}
Note that our expression properly recovers the  limits discussed in the preceding subsections, Eq.~(\ref{sigmaxxc}) and $q_\mathrm{c}=0$ for large $d$, and Eqs.~(\ref{qcsmalld}) and (\ref{sigmacsmalld}) for small $d$. }

Also, note that when the stress is above the critical stress, there is a bifurcation and there are two modes of buckling, a short wavelength mode and a long wavelength mode. The wavelength of the short-wavelength mode changes slowly with stress. On the other hand,  the wavelength of the long-wavelength mode changes rapidly with stress, which sometimes makes it hard to observe, although it has been seen experimentally~\cite{Shield_etal1994,Efimenko_etal2005,Sun_etal2011} and in computations~\cite{Efimenko_etal2005,Sun_etal2011} of wrinkling on isotropic substrates.

\subsection{\trpc{The limit of vanishing anisotropy}}\label{smallr11}
\trpc{The results described above do not approach the isotropic case when $r\to1$ since the transition from nematic to the isotropic phase of the elastomer is discontinuous. However, if the ideal nature of the elastomer is spoiled by fluctuations in the composition of the chains, then soft elasticity is replaced by semi-soft elasticity~\cite{WarnerTerentjev}.}
\trpc{When semi-softness is accounted for and $r$ is near unity,} the second term on the right-hand side of Eq.~\eqref{d1} \trpc{dominates the expression for the  modulus $D_1$ coupling director rotation and the local rotation of the polymer matrix.}  
Frank elasticity can be ignored in this case since $K$ goes as $(r-1)^2$, \trpc{and therefore t}he decay length of the deformations will be of order $1/q$, \trpc{just as for an isotropic substrate}. 
\trpc{S}ince $qd\gg 1$, we can assume that the substrate has infinite thickness. Using the boundary conditions (\ref{psiasympt}-\ref{thetaasympt}) we find \trpc{to leading order in $(r-1)$}
\begin{equation}
	\sigma_n\simeq2 \trpc{C_1} \hs{u^\mathrm{f}_z} \mu  q, \label{semisoft1}
\end{equation}
\trpc{where}
\begin{equation}
	C_1=q\left[ 1-\frac{  (r-1) (r+1)^2}{8 a r^2}\right], \label{C1}
\end{equation}
\trpc{and} $a=a_1+(r-1)/r$. 
\trpc{Note that $C_1=1$ when $r-1$. When $r$ is close to unity, the critical stress and wavenumber are the same as in the case of an isotropic elastomer substrate, except the modulus $E$ is replaced by $C_1 E$ (see Fig.~\ref{sigma_q}).}

\section{Discussion}

\trpc{Table~\ref{tablescales} summarizes the critical compressive stress and critical wavenumber for the different cases we consider in this article. In all cases, the substrate is incompressible. To see what case best describes the experiments of Agrawal and collaborators~\cite{aditya2012SoftMatter}, we calculate the penetration depth $l_\mathrm{p}$ using Eq.~(\ref{eqn:lp}) for the experimental parameters quoted above ($\mu=0.24\,$MPa, $E_\mathrm{f}=3.5\,$GPa, $\kappa=10^{-11}\,$N) assuming the film is incompressible ($\nu_\mathrm{f}=1/2$) and using a typical value of the step-length anisotropy parameter $r=1.7$. For the thinnest films considered, $h=30\,$nm, we find that the penetration depth corresponding to $q_\mathrm{c}$ [see Eq.(~\ref{fullqc})] is $l_\mathrm{p}\approx 1.3\,$mm, which is about three times the substrate thickness used, $d=360\,\mu$m. When $h=100$\,nm, the penetration depth is about 25 times the substrate thickness.}
Therefore, for all but the thinnest films we may use the small $d$ limit (last row of Table~\ref{tablescales}) for the critical wavelength  and stress that develops upon heating the system,
\begin{eqnarray}
q_\mathrm{c}h&=&2\left(\frac{E}{\bar{E}_\mathrm{f}}\right)^{1/4}\left(\frac{h}{d}\right)^{1/4}\\
-\sigma_{\mathrm{c}}^\mathrm{f}&=&\frac{2}{3}\sqrt{E\bar{E}_\mathrm{f}}\left(\frac{h}{d}\right)^{1/2}.
\end{eqnarray}
\trpc{This critical stress is much less than the critical stress for an isotropic elastomer with the same material parameters since $h\ll d$. Note however that cooling the system leads to wrinkles with crests parallel to the nematic directors. In this situation, the elastomer behaves as an isotropic rubber when responding to the wrinkled film; therefore the critical stress is predicted to be highly anisotropic, with different values for heating and cooling.}

\trpc{We can also calculate the typical compressive stress used in the experiments. The different rates of expansion $\alpha_\mathrm{f}$ and $\alpha$ of the film and the substrate, respectively, lead to a compressive stress of}
\begin{equation}
	\sigma^\mathrm{f}_{xx}= -\dfrac{E_\mathrm{f} (\alpha_\mathrm{f}-\alpha )\Delta T}{1-\nu_\mathrm{f}}.
\end{equation}
The coefficient of thermal expansion  for  the nematic elastomer substrate is calculated from ~\cite{WarnerTerentjev}
\begin{equation}
\alpha=\frac{\lambda _m-1}{\Delta T},
\end{equation}
where $\lambda _m=\sqrt[3]{r/r_0}$, and $r_0$ and $r$ are the initial and final values of the step length anisotropic ratio. The step length anisotropic ratios $r_0$ and $r$ can be calculated from the values of $\ell_{\|}$ and $\ell_{\perp}$ extracted from the Fig. (2) of  \trp{Agrawal} \textit{et al}. Values of  $r_0$ and $r$ are 1.75 and 1.68  for the wavelength vs. thickness plot in \trp{Agrawal} \textit{et al.}, which give $\alpha=-9.0\times10^{-4}$ K$^{-1}$. The coefficient of thermal expansion for the thin film  is  $\alpha_\mathrm{f}=7 \times 10^{-5}$ K$^{-1}$. \trpc{Using these values leads to a compressive stress of $-\sigma^\mathrm{f}_{xx}=$9.32$\times10^7$\,Pa. Thus, in the experiments~\cite{aditya2012SoftMatter} the typical compressive stresses are large compared to the critical stress, and well in the regime where the dependence of the stress on the wrinkling wavenumber is the same for the isotropic and nematic elastomers (see Fig.~\ref{sigma_q}). } 

\section{Conclusion}

\trpc{We have studied the wrinkling of a thin film on a soft nematic elastomer substrate, and found that when the wrinkles are perpendicular to the director in the elastomer, the critical compressive stress is much less that that for an isotropic elastomer.  As a consequence of the soft elasticity of nematic elastomers, the deformation accompanying the wrinkles penetrates deep into the elastomer. This penetration depth is so long that for film thicknesses greater that a few tens of nanometers, a substrate of a few hundred microns can be considered thin enough that the deformation penetrates all the way though the substrate. Our work has been limited to small amplitude wrinkles and the calculation of the critical stress and wrinkling wavenumber. A natural extension would be to study wrinkling patterns for large deformation, and also to allow the directors to tilt out of the plane perpendicular to the wrinkles. It would be interesting to study the effect of soft elasticity on related large-amplitude phenomena such as the formation of  multi-mode wrinkle clusters and creases~\cite{Sun_etal2011}.  }





\begin{acknowledgments} 
This work was supported in part by National Science Foundation Grant MRSEC-1420382. We are grateful to \trpa{Benny Davidovitch, Marcelo Dias}, KS Kim, \trpa{Ski Krieger, Narayanan Menon, and} Mark Warner for helpful discussions.
\end{acknowledgments}


%

\end{document}